\documentclass[12pt,onecolumn,final]{IEEEtran}

\usepackage{cite}
\ifCLASSINFOpdf
  \usepackage{graphicx,epstopdf}
  \graphicspath{{../Figures/}}
  \DeclareGraphicsExtensions{.eps}
\else
  \usepackage[dvips]{graphicx}
\fi
\usepackage[cmex10]{amsmath}
\usepackage[tight,footnotesize]{subfigure}
\usepackage{amssymb}
\usepackage{url}

\begin{document}
\title{Linear Precoding for Broadcast Channels with Confidential Messages under Transmit-Side Channel Correlation}
\author{\normalsize\authorblockN{{Giovanni~Geraci$^{1,2}$,~Azzam~Y.~Al-nahari$^{3,4}$,~Jinhong~Yuan$^1$~and~Iain~B.~Collings$^2$}}\\
\small\authorblockA{$^1$School of Electrical Engineering \& Telecommunications,
The University of New South Wales, \textsc{Australia} }\\
\authorblockA{$^2$Wireless and Networking Technologies Laboratory, CSIRO ICT Centre, Sydney, \textsc{Australia}}\\
\authorblockA{$^3$Department of Electrical and Information Technology, Lund University, \textsc{Sweden}}\\
\authorblockA{$^4$Department of Electronics Engineering, Ibb University, \textsc{Yemen}}\\
}
\maketitle
\begin{abstract}
In this paper, we analyze the performance of regularized channel inversion (RCI) precoding in multiple-input single-output (MISO) broadcast channels with confidential messages under transmit-side channel correlation. We derive a deterministic equivalent for the achievable per-user secrecy rate which is almost surely exact as the number of transmit antennas and the number of users grow to infinity in a fixed ratio, and we determine the optimal regularization parameter that maximizes the secrecy rate. Furthermore, we obtain deterministic equivalents for the secrecy rates achievable by: (i) zero forcing precoding and (ii) single user beamforming. The accuracy of our analysis is validated by simulations of finite-size systems.
\end{abstract}

\IEEEpeerreviewmaketitle
\thispagestyle{empty}
\newpage
\section{Introduction}
\setcounter{page}{1}
Wireless multiuser communications are very susceptible to eavesdropping, and the transmitted information has traditionally been secured at the network layer with cryptographic schemes.  However, this raises issues like key distribution and management, and high computational complexity. Moreover, these schemes rely on the unproven assumption that certain mathematical functions are hard to invert. Therefore, it is of critical importance to tackle security at the physical layer.

Physical layer security was introduced in \cite{Wyner75,Csiszar78} to exploit the randomness inherent in noisy channels. The secrecy capacity region of a broadcast channel with confidential messages (BCC), where the intended users can act maliciously as eavesdroppers, was studied in \cite{Liu08,Liu13} among others, but only for the case of two users. For a larger BCC with any number of malicious users, it was shown in \cite{Geraci12,GeraciJSAC} that regularized channel inversion (RCI) precoding can achieve large secrecy sum-rates with low-complexity implementation, if the channels are independent and identically distributed (i.i.d). However in practice, channel correlation is present due to local scatterers around the base station and insufficient separation between antennas.
The large-system performance of RCI precoding in the broadcast channel (BC) without secrecy requirements under channel correlation was analyzed in \cite{Wagner12,Muharar}, where deterministic approximations were obtained for the sum-rates.

In this paper, we analyze the secrecy sum-rate achievable by RCI precoding in the multiple-input single-output (MISO) BCC under transmit-side channel correlation. This work directly extends some of the analysis in \cite{Wagner12,Muharar} by requiring the transmitted messages to be kept confidential. Furthermore, this work generalizes the results in \cite{Geraci12,GeraciJSAC}, where no channel correlation was assumed. In the analysis, we provide a deterministic approximation for the large-system secrecy rates achievable by RCI precoding under transmit-side correlation, and we derive the optimal regularization parameter of the RCI precoder that maximizes the large-system secrecy rates. Then we obtain deterministic equivalents for the secrecy rates achievable by: (i) zero forcing precoding and (ii) single user beamforming. Simulations demonstrate that the analysis is accurate even when applied to finite-size systems.

\section{MISO BCC under Transmit-Side Channel Correlation}

\subsection{System Model}

We consider the downlink of a narrowband multi-user MISO system, consisting of a base station (BS) with $M$ antennas which simultaneously transmits $K$ independent confidential messages to $K$ spatially dispersed single-antenna users. In this model, transmission takes place over a block fading channel, and the transmitted signal is $\mathbf{x} = \left[x_1,\ldots,x_M \right]^{T} \in \mathbb{C}^{M \times 1}$. The received signal at user $k$ is given by
\begin{equation}
y_k=\sum_{j=1}^{M} h_{k,j}x_{j}+n_{k}
\label{eqn:MIMO_scalar}
\end{equation}
where $h_{k,j}$ is the channel between the $j$-th transmit antenna element and the $k$-th user, and $n_{k}$ is the noise seen at the $k$-th receiver. 
The corresponding vector equation is
\begin{equation}
\mathbf{y}=\mathbf{Hx}+\mathbf{n}
\label{eqn:MIMO_vector}
\end{equation}
where $\mathbf{H} \in \mathbb{C}^{K \times M}$ is the channel matrix with entries $h_{k,j}$, $\mathbf{y} = \left[ y_1, \ldots, y_K \right] ^{T} \in \mathbb{C}^{K \times 1}$ and $\mathbf{n} = \left[ n_1, \ldots, n_K \right] ^{T} \in \mathbb{C}^{K \times 1}$. We assume that $\mathbb{E}[ \mathbf{nn}^{\dagger} ] =\sigma^{2} \mathbf{I}$, define the SNR $\rho \triangleq 1/ \sigma ^2$, and impose the long-term power constraint $\mathbb{E}[ \left\| \mathbf{x} \right\|^{2} ] =1$. The transmitted signal $\mathbf{x}$ is obtained at the BS by performing a linear processing on the vector of confidential messages $\mathbf{u} = \left[u_1,\ldots,u_K \right]^{T} \in \mathbb{C}^{K \times 1}$, whose entries are chosen independently, satisfying $\mathbb{E}[ \left|u_k\right|^2 ] =1$. We assume homogeneous users, i.e. each user experiences the same received signal power on average, thus the model assumes that their distances from the transmitter are similar.

We consider correlation amongst transmit antenna elements, and for tractability in the analysis, we assume a separable (or Kronecker) correlation model \cite{Muharar,Al-naffouri}. Hence the channel matrix can be written as
\begin{equation}
\mathbf{H} = \mathbf{\tilde{H}R}^{\frac{1}{2}}
\end{equation}
where $\mathbf{\tilde{H}}\in \mathbb{C}^{K \times M}$ contains i.i.d. circularly-symmetric complex Gaussian random variables, with zero mean and unit variance, and where $\mathbf{R}\in \mathbb{C}^{M \times M}$ is the non-singular transmit correlation matrix, given by $\mathbf{R} = \mathbb{E}[ \mathbf{H}^{\dagger}\mathbf{H} ]$. We assume that each user experiences the same transmit correlation. The channel vector for the $k$-th user is $\mathbf{h}_k^{\dagger} = \mathbf{\tilde{h}}_k^{\dagger} \mathbf{R}^{\frac{1}{2}} \in \mathbb{C}^{M \times 1}$.

It is required that the BS securely transmits each message $u_k$, ensuring that the unintended users receive no information. We assume that for each message $u_k$ the $K-1$ unintended users can cooperate to jointly eavesdrop on the intended user $k$, and form an equivalent single eavesdropper with $K-1$ receive antennas, denoted by $\widetilde{k}$. Although this is impossible in practice, such worst-case assumption is required to guarantee the secrecy of all messages, since the behavior of the users cannot be predicted. The secrecy sum-rate is given by  
\begin{equation}
R_s \triangleq \sum_{k=1}^{K} {R_{k}},
\end{equation}
where each secrecy rate $R_{k}$ is defined as the rate at which the BS communicates reliably to user $k$ without allowing $\widetilde{k}$ to obtain any information about $u_k$.
\subsection{Secrecy Sum-Rates with RCI Precoding}

In RCI precoding, the transmitted vector is obtained as $\mathbf{x} = \mathbf{Wu}$, with the precoding matrix $\mathbf{W} \in \mathbb{C}^{M \times K} $ given by \cite{Peel05}
\begin{equation}
\mathbf{W} = \left[ \mathbf{w}_1, \cdots, \mathbf{w}_K \right] = \frac{1}{\sqrt{\gamma}} \mathbf{H}^{\dagger} \left( \mathbf{H H}^{\dagger} + \xi M \mathbf{I}_K \right) ^{-1} \mathbf{u},
\label{eqn:RCI_precoder}
\end{equation}
and
\begin{equation}
\gamma = \textrm{Tr} \{ \mathbf{H}^{\dagger} \mathbf{H} \left( \mathbf{H}^{\dagger} \mathbf{H} + \xi M \mathbf{I}_M \right) ^{-2} \}
\label{eqn:gamma}
\end{equation}
is a long-term power normalization constant. The function of the real regularization parameter $\xi\geq0$ is to achieve a tradeoff between the signal power at the intended user and the interference and information leakage at the unintended users for each message.

For the multi-user MISO system with malicious users, a secrecy sum-rate $R_s$ achievable by linear precoding was found in \cite{Geraci12} by considering that each user $k$, along with its own eavesdropper $\widetilde{k}$ and the transmitter, forms an equivalent multi-input, single-output, multi-eavesdropper (MISOME) wiretap channel \cite{Khisti10I}. The transmitter, the intended receiver and the eavesdropper of each MISOME wiretap channel are equipped with $M$, $1$ and $K-1$ virtual antennas, respectively.
Due to the simultaneous transmission of the $K$ messages, the receiver $k$ experiences noise and interference from all the messages $u_j$ with $j \neq k$. Since the malicious users can cooperate, interference cancellation can be performed at the receiver $\widetilde{k}$, which does not see any undesired signal term apart from the received noise, and $R_s$ represents a lower-bound on the achievable secrecy sum-rate.

For a given channel $\mathbf{H}$, let $\mathbf{H}_{\widetilde{k}}$ be the matrix obtained from $\mathbf{H}$ by eliminating the $k$-th row. Then a lower-bound on the secrecy sum-rate achievable by RCI precoding is \cite{Geraci12}
\begin{equation}
R_{s} = \sum_{k=1}^{K} { R_{k}} = \sum_{k=1}^{K} { \left\{ \log_2 \frac{1 + \frac {\rho \left| \mathbf{h}_k^{\dagger} \mathbf{w}_k \right| ^2} {\gamma + \rho \sum_{j \neq k} {\left| \mathbf{h}_k^{\dagger} \mathbf{w}_j \right| ^2} }}{1 +  \frac{\rho}{\gamma } \left\| \mathbf{H}_{\widetilde{k}} \mathbf{w}_{k} \right\| ^2} \right\}^+ }.
\label{eqn:Rs_RCI}
\end{equation}
In the following, we refer to (\ref{eqn:Rs_RCI}) as the secrecy sum-rate.

\section{Large System Analysis}

In this section, we analyze the performance of the RCI precoder under transmit-side channel correlation in the large-system regime, where both the number of receivers $K$ and the number of transmit antennas $M$ approach infinity, with their ratio $\beta = K/M$ being held constant and finite.

\subsection{Large-System Secrecy Sum-Rates}

The following theorem provides a deterministic equivalent for the per-user secrecy rates $R_k$, which is almost surely exact as $M \rightarrow \infty$.

\newtheorem{Theorem}{Theorem}
\begin{Theorem}
Let $\rho>0$, $\beta>0$, and $\xi>0$. Then for all $k$
\begin{equation}
\left| R_k - R^{\circ} \right| \stackrel{\textrm{a.s.}}{\longrightarrow} 0, \quad \textrm{as} \quad M \rightarrow \infty.
\label{eqn:Theorem1}
\end{equation}
$R^{\circ}$ is the large-system secrecy rate, given by
\begin{equation}
R^{\circ} = \left\{ \log_2 \! \! \frac{1\!+\!\eta \frac{\rho E_{22} + \frac{\xi\rho}{\beta}(1+\eta)^2 E_{12}}{\rho E_{22}+(1+\eta)^2E_{12}}}{1+\frac{\rho E_{22}}{(1+\eta)^2 E_{12}}} \! \right\}^+\!,
\label{eqn:Rs_large_system}
\end{equation}
where
\begin{equation}
\eta = \mathbb{E} \left[ \frac{\texttt{T}(1+\eta)}{\xi(1+\eta)+\beta\texttt{T}} \right], \quad \textrm{and} \quad
E_{ij} = \mathbb{E} \left[ \frac{\texttt{T}^i}{\left(\xi(1+\eta)+\beta\texttt{T}\right)^j} \right].
\label{eqn:zeta_E}
\end{equation}
The expectations in (\ref{eqn:zeta_E}) are taken over the random variable $\texttt{T}$, whose distribution function $\Lambda(t)$ is the limiting eigenvalue distribution of the correlation matrix $\mathbf{R}$.
\end{Theorem}
\begin{IEEEproof}
See Appendix A for a sketch of the proof.
\end{IEEEproof}
We note that (\ref{eqn:Rs_large_system}) is a non-random function of $\rho$, $\beta$, and $\xi$. 

\newtheorem{Corollary}{Corollary}
\begin{Corollary}
In the case of no channel correlation, $R^{\circ}$ reduces to the value obtained in \cite{GeraciJSAC}, given by
\begin{equation}
\widetilde{R}^{\circ} = \left\{ \log_2 \frac{1+
g\left( \beta,\xi \right)
\frac{\rho + \frac{\rho\xi}{\beta} \left[ 1 + g\left( \beta,\xi \right) \right] ^2 
}{\rho + \left[ 1 + g\left( \beta,\xi \right) \right] ^2}}
{1+\frac{\rho}{ \left[ 1+g\left( \beta,\xi \right) \right] ^2}} \right\}^+
\label{eqn:Rs_nocorr},
\end{equation}
where $g \left( \beta,\xi \right) \triangleq \frac{1}{2} \left[ \sqrt{ \frac{\left(1\!-\!\beta \right)^2}{\xi^2} \! + \! \frac{2\left(1\!+\!\beta\right)}{\xi} \! + \! 1} +  \frac{1\!-\!\beta}{\xi} \! - \! 1 \right]$.
\end{Corollary}
\begin{IEEEproof}
See Appendix B for a sketch of the proof.
\end{IEEEproof}

\subsection{Selection of the Optimal Regularization Parameter}

The value of $\xi$ affects the asymptotic secrecy rate $R^{\circ}$ in (\ref{eqn:Rs_large_system}). We now study the optimal regularization parameter $\xi^{\circ\star}$ that maximizes $R^{\circ}$ under transmit-side channel correlation.

\begin{Theorem}
The regularization parameter $\xi^{\circ\star}$ that maximizes the secrecy rate $R^{\circ}$ under transmit-side channel correlation can be obtained by solving the fixed-point equation
\begin{equation}
  \xi = \beta \frac{(\eta^2-1)E_{12}-\rho E_{22}}{2\rho\eta(1+\eta)E_{12}}.
\label{eqn:xi_opt}
\end{equation}
\end{Theorem}
\begin{IEEEproof}
See Appendix C for a sketch of the proof.
\end{IEEEproof}

We note that $\eta$, $E_{12}$, and $E_{22}$ in (\ref{eqn:xi_opt}) depend on $\xi$. It is easy to verify that the value of the regularization parameter $\xi^{\circ\star}$ that maximizes $R^{\circ}$ differs from the value $\beta/\rho$ that maximizes the rates without secrecy requirements \cite{Muharar}. Moreover in the presence of secrecy requirements, the channel correlation affects the optimal regularization parameter.

\subsection{Comparison with Other Precoding Schemes}

\subsubsection{Zero Forcing}

As the regularization parameter $\xi \rightarrow 0$, we have the zero forcing (ZF) precoder. The aim of the ZF precoder is to cancel all the interference and information leakage. With ZF precoding the transmitted signal is
\begin{equation}
  \mathbf{x} = \left\{ 
  \begin{array}{l l l}
\frac{1}{\sqrt{\gamma_0}} \mathbf{H}^{\dagger} \left( \mathbf{H H}^{\dagger} \right) ^{-1} \mathbf{u} \quad \text{for $\beta \leq 1$}\!\\
\frac{1}{\sqrt{\gamma_0}} \left( \mathbf{H}^{\dagger} \mathbf{H} \right) ^{-1} \mathbf{H}^{\dagger} \mathbf{u} \quad \text{for $\beta > 1$}\!
  \end{array} \right.
\label{eqn:ZF_precoder}
\end{equation}
where $\gamma_0$ is the power normalization constant.
It can be shown that in the large-system regime, the secrecy rates $R_{0}^{\circ}$ achievable by ZF precoding under transmit-side channel correlation are given by
\begin{equation}
  R_{0}^{\circ} = \left\{ 
  \begin{array}{l l}
\log_2 \left( 1 + \frac{\rho}{\chi} \right) &\quad \text{for $\beta \leq 1$}\!\\
\left\{ \log_2 \frac{1+\frac{\rho}{\rho(\beta-1)+\beta^2\gamma_0^{\circ}}}{1 + \frac{\rho(\beta-1)}{\gamma_0^{\circ}\beta^2}}\right\}^+  &\quad \text{for $\beta > 1$}\!
  \end{array} \right.
\label{eqn:Rs_ZF}
\end{equation}
with
\begin{equation}
\chi = \frac{\beta}{\int{\frac{t d\Lambda(t)}{1+\chi t}}} \quad \textrm{and} \quad \gamma_0^{\circ} = \frac{\beta-1}{\int{\frac{d\Lambda(t)}{t}}}.
\end{equation}

\subsubsection{Single User Beamforming}

As the regularization parameter $\xi \rightarrow \infty$, we have the single user beamformer (SUB). Here, the transmitter beamforms
in a direction such as to maximize the signal strenght
of each user, without taking into account the interference it
creates and the amount of resulting information leakage. The SUB is given by
\begin{equation}
\mathbf{x} = \frac{1}{\sqrt{\gamma_{\infty}}} \mathbf{H}^{\dagger} \mathbf{u},
\end{equation}
where $\gamma_{\infty}$ is the power normalization constant.
It can be shown that the secrecy rates $R_{\infty}^{\circ}$ achievable by SUB in the large-system regime and under transmit-side channel correlation are
\begin{equation}
R_{\infty}^{\circ} = \left\{ \log_2 \frac{1 + \frac{\rho \mathbb{E}^2[\texttt{T}]}{\beta(\rho\mathbb{E}[\texttt{T}^2]+\mathbb{E}[\texttt{T}])}}{1+\frac{\rho\mathbb{E}[\texttt{T}^2]}{\mathbb{E}[\texttt{T}]}}\right\}^+.
\label{eqn:Rs_SUB}
\end{equation}

We note that when $\beta \geq 1$, it is $R_{\infty}^{\circ}=0$ $ \forall \rho$. When $\beta <1$, there is always a value of $\rho$ beyond which $R_{\infty}^{\circ}=0$. Such poor performance is due to the intended user suffering from a large amount of interference, while the eavesdroppers may cancel the interference by cooperating.
\section{Numerical Results}

Calculating $R^{\circ}$ in (\ref{eqn:Rs_large_system}) requires the limiting eigenvalue distribution of the correlation matrix $\mathbf{R}$. In this section, we consider the Toeplitz-exponential model \cite{Gray05}, where $\mathbf{R}$ has a Toeplitz structure and its entries follow the distribution $r_{ij} = \nu^{|i-j|}$, governed by the correlation coefficient $\nu \in [0,1]$. It can be shown that under Toeplitz-exponential correlation, the large-system secrecy rates $R^{\circ}$ reduce to
\begin{equation}
R_{TE}^{\circ} = \left\{ \log_2 \! \! \frac{1\!+\!\eta \frac{\rho c + \frac{\xi\rho}{\beta}(1+\eta)^2}{\rho c+(1+\eta)^2}}{1+\frac{\rho c}{(1+\eta)^2}} \! \right\}^+\!,
\label{eqn:Rs_large_system_TE}
\end{equation}
with $c=\frac{\xi(1+\eta)(1+\nu^2)+\beta(1-\nu^2)}{\xi(1+\eta)(1-\nu^2)+\beta(1+\nu^2)}$.

\begin{figure}[t]
\centering
\includegraphics[width=3.75in]{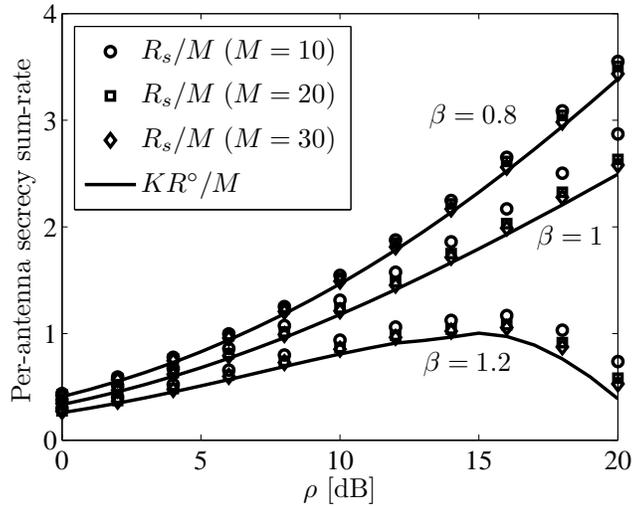}
\caption{Comparison between the simulated ergodic per-antenna secrecy sum-rate $R_s/M$ from (\ref{eqn:Rs_RCI}) and the large-system approximation $K R^{\circ}/M$ from (\ref{eqn:Rs_large_system}), for $\nu=0.5$ and various values of $\beta$.}
\label{fig:Rs_analysis_vs_sim}
\end{figure}

Fig. \ref{fig:Rs_analysis_vs_sim} compares the per-antenna secrecy sum-rate approximation $K R^{\circ}/M$ under transmit-side correlation to the simulated ergodic per-antenna secrecy sum-rate $R_s/M$ with a finite number of users, for $\nu=0.5$ and three values of $\beta$. The values of $R^{\circ}$ and $R_s$ were obtained by (\ref{eqn:Rs_large_system}) and (\ref{eqn:Rs_RCI}), respectively, using $\xi = \xi^{\circ\star}$ obtained as a positive solution to (\ref{eqn:xi_opt}). We note that the accuracy of the analysis decreases with $\rho$.
The loss of accuracy is due to the limitations of the tools used to derive the deterministic equivalents \cite{Wagner12}. One can increase the accuracy at any given $\rho<\infty$ by increasing the dimension $M$.

\begin{figure}[t]
\centering
\includegraphics[width=3.75in]{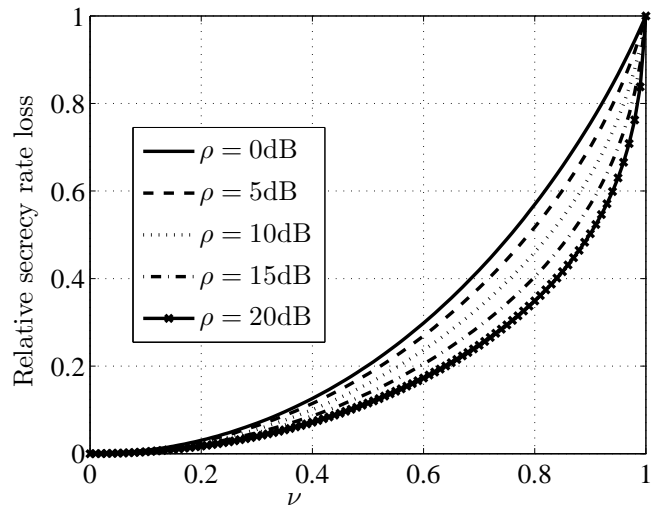}
\caption{Relative secrecy rate loss $(\widetilde{R}^{\circ}-R^{\circ})/\widetilde{R}^{\circ}$ as a function of the correlation coefficient $\nu$, for $\beta=0.8$.}
\label{fig:loss}
\end{figure}

Fig. \ref{fig:loss} shows the relative secrecy rate loss $(\widetilde{R}^{\circ}-R^{\circ})/\widetilde{R}^{\circ}$ due to channel correlation, with $R^{\circ}$ and $\widetilde{R}^{\circ}$ from (\ref{eqn:Rs_large_system}) and (\ref{eqn:Rs_nocorr}), respectively. We observe how low-to-moderate correlation , i.e. $\nu<0.4$, does not affect the secrecy sum-rate excessively, i.e. the loss is less than 10 percent. However, higher correlation can significantly degrade the performance of the RCI precoder, especially at low SNR.

\begin{figure}[t]
\centering
\includegraphics[width=3.75in]{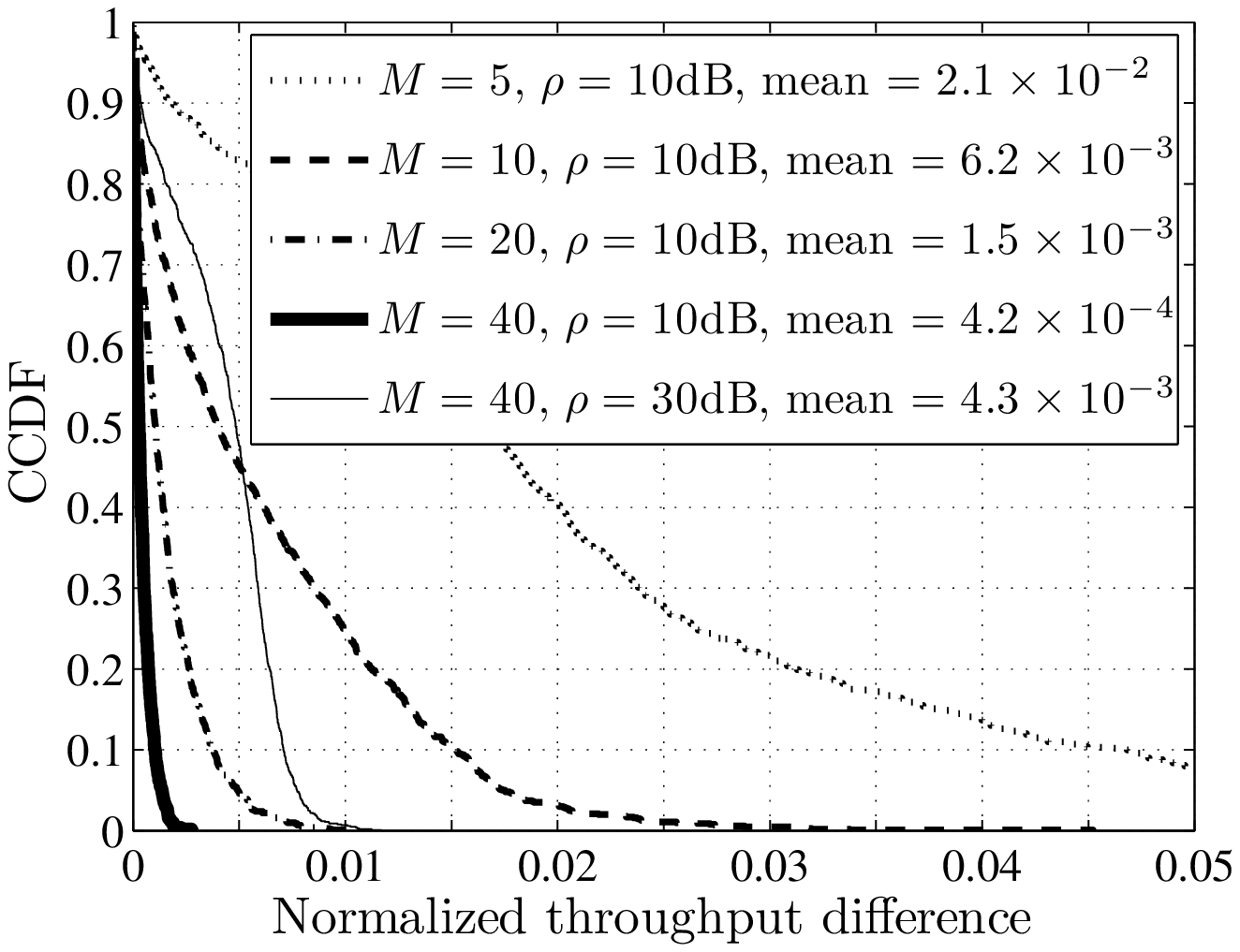}
\caption{CCDF of the normalized secrecy sum-rate difference between using: (i) $\xi^{\circ\star}$ obtained from (\ref{eqn:xi_opt}) and (ii) $\xi^{\star}$ obtained by bi-sectional search for every channel realization, for $\beta=1$, and $\nu=0.5$.}
\label{fig:CCDF}
\end{figure}

Fig. \ref{fig:CCDF} shows that using the large-system regularization
parameter $\xi^{\circ\star}$ from (\ref{eqn:xi_opt}) does not cause a significant loss in the secrecy sum-rate compared to using a finite-system regularization parameter $\xi=\xi^{\star}$ optimized by bi-sectional search for each channel realization. The complementary cumulative distribution function (CCDF) of the normalized secrecy sum-rate difference $(R_s(\xi^{\star})-R_s(\xi^{\circ\star}))/R_s(\xi^{\star})$ between using $\xi^{\circ\star}$ and $\xi^{\star}$ depends on the values of SNR and $\beta$, where large SNR and $\beta=1$ represent the worst case. The CCDFs in Fig. \ref{fig:CCDF} were obtained for $\beta=1$, and $\nu=0.5$. Since the average normalized secrecy sum-rate difference is two percent or less for all values of $M$, the large-system regularization parameter $\xi^{\circ\star}$ from (\ref{eqn:xi_opt}) may be used instead of the finite-system regularization parameter with only a small loss of performance. Such choice avoids the computation of $\xi^{\star}$ for each channel realization.

\section{Conclusion}
In this paper, we considered regularized channel inversion (RCI) precoding for the MISO broadcast channel with confidential messages, under transmit-side channel correlation. We first analyzed the RCI precoder in the large-system regime, and derived a deterministic equivalent for the achievable per-user secrecy rate, as well as the optimal regularization parameter that maximizes the deterministic equivalent of the per-user secrecy rate. Then we obtained deterministic equivalents for the secrecy rates achievable by: (i) zero forcing precoding and (ii) single user beamforming. The large-system approximations prove to be accurate even for finite-size systems. Simulations showed that low-to-moderate correlation only partially affects the secrecy rates. However, high correlation degrades the performance of the RCI precoder, especially at low SNR.
\section*{Appendix}

\subsection{Sketch of Proof of Theorem 1}

By introducing the quantities \cite{NguyenGCOM08}
\begin{equation}
A_k = \mathbf{\tilde{h}}_k^{\dagger} \mathbf{Q}_k \mathbf{\tilde{h}}_k, \enspace B_k \! = \! \mathbf{\tilde{h}}_k^{\dagger} \! \mathbf{Q}_k \! \mathbf{\tilde{H}}_{\widetilde{k}}^{\dagger} \mathbf{\tilde{H}}_{\widetilde{k}} \mathbf{Q}_k \! \mathbf{\tilde{h}}_k, \enspace \textrm{and} \enspace \gamma = \textrm{Tr} \left\{ \mathbf{Q}_k \mathbf{\tilde{H}R}^{-1} \mathbf{\tilde{H}}^{\dagger} \mathbf{Q}_k \right\},
\label{eqn:Ak_Bk_gamma_bis}
\end{equation}
with $\mathbf{Q}_k=\left( \mathbf{\tilde{H}}_{\widetilde{k}}^{\dagger} \mathbf{\tilde{H}}_{\widetilde{k}} + M \xi \mathbf{R}^{-1} \right) ^{-1}$, it is possible to express the secrecy rates as
\begin{equation}
R_{k} = { \left\{ \log_2 \frac{ 1 + \frac{\rho A_k^2}{\rho B_k + \gamma \left( 1 + A_k \right)^2} } {1 + \frac{\rho B_k}{\gamma \left( 1 + A_k \right)^2 }  }
\right\}^+ }.
\label{eqn:Rs_RCI_bis}
\end{equation}

As $M,K \rightarrow \infty$, the quantities in (\ref{eqn:Ak_Bk_gamma_bis}) respectively converge almost surely to \cite{Muharar}
\begin{equation}
A_k^{\circ} = \eta, \quad
B_k^{\circ} = \frac{\beta E_{22}}{1-\beta E_{22}}, \quad \textrm{and} \quad
\gamma^{\circ} =  -\frac{\beta}{(1+\eta)^2}\frac{\partial \eta}{\partial \xi},
\label{eqn:A_B_gamma_o}
\end{equation}
where $\frac{\partial \eta}{\partial \xi} = \frac{-(1+\eta)^2E_{12}}{1-\beta E_{22}}$ and $\eta$, $E_{12}$ and $E_{22}$ are defined in (\ref{eqn:zeta_E}). Then (\ref{eqn:Rs_large_system}) follows from (\ref{eqn:Rs_RCI_bis}), (\ref{eqn:A_B_gamma_o}), by applying the continuous mapping theorem, the Markov inequality, and the Borel-Cantelli lemma \cite{CouilletBook}.

\subsection{Sketch of Proof of Corollary 1}

When no channel correlation is present, we have $\mathbf{R}=\mathbf{I}$ and $\Lambda(t)=\delta(t-1)$, which implies $\eta = g$ and $E_{12} = E_{22} = [\xi(1+g)+\beta]^{-2}$. Therefore $A_k^{\circ} = g$, $B_k^{\circ} = \frac{\beta g}{\xi(1+g)^2+\beta}$, and $\gamma^{\circ} = \frac{\beta g}{\xi(1+g)^2+\beta}$, and (\ref{eqn:Rs_nocorr}) follows.

\subsection{Sketch of Proof of Theorem 2}

We obtain eq. (\ref{eqn:xi_opt}) by defining $\chi \!=\! (1\!+\!\eta)^2 E_{12}$, $Z \!=\! (\beta E_{22} \!+\! \xi \chi)(\rho E_{22} \!+\! \chi)$, $\Psi \!=\! \frac{\beta E_{22} + \xi(1+\eta)^2 E_{12}}{\rho E_{22}+(1+\eta)^2E_{12}}$, $\Phi \!=\! \frac{\rho^2 \Psi \eta \Lambda}{\beta Z}$, and $\Lambda \!=\! \frac{2 \eta' (1\!+\!\eta) E_{12} E_{22} + (1\!+\!\eta)^2 (E_{12}' E_{22} \!-\! E_{22}' E_{12})}{\eta'}$ \cite{Muharar}, with $E_{12}' E_{22} - E_{22}' E_{12} = -2 \beta (1+ \eta + \xi \eta') ( E_{13}E_{33}-E^2_{23})$, and solving
\begin{equation}
\frac{\partial R^{\circ}}{\partial \xi} = \frac{\eta'}{2^{R_s^{\circ}} \rho \chi^2 \beta (1\!+\!\frac{\rho E_{22}}{\chi})^2 \log 2}
[\Phi (\rho \xi \!-\!\beta)(\chi^2\!+\!\rho \chi E_{22})\beta + \rho^2 \Lambda (\beta+\rho \eta \Psi)]=0.
\label{eqn:derivative}
\end{equation}
\ifCLASSOPTIONcaptionsoff
  \newpage
\fi
\bibliographystyle{IEEEtran}

\vspace{-0.2cm}
\bibliography{IEEEabrv,Bib_Giovanni}
\end{document}